# A POWERFUL OPTIMIZATION APPROACH FOR THE MULTI CHANNEL DISSEMINATION NETWORKS

Ahmad Saad Al-Mogren

Department of Computer Science, King Saud University, Riyadh, Saudi Arabia
`almogren@hotmail.com`

## ABSTRACT

*In the wireless environment, dissemination techniques may improve data access for the users. In this paper, we show a description of dissemination architecture that fits the overall telecommunication network. This architecture is designed to provide efficient data access and power saving for the mobile units. A concurrency control approach, MCD, is suggested for data consistency and conflict checking. A performance study shows that the power consumption, space overhead, and response time associated with MCD is far less than other previous techniques.*

## 1. INTRODUCTION

Advances in technology facilitate the use of data dissemination (i.e. push-based) applications. These applications involve the dissemination of data, such as weather information and stock quotes, to a large set of listening clients. Data dissemination applications possess some unique characteristic, such as similar user profiles and asymmetric data flow from servers to clients. In fact, there are many advantages to the data dissemination approach. Firstly, many users can access data simultaneously. This increases not only the parallelism but also more effectively utilizes the limited capacity of the wireless channel and, therefore, improves the performance of the mobile system. Secondly, data dissemination applications send data to the listening users through pushing data rather than pulling, and therefore, avoiding the slow uplink channel speed.

There are a lot of research issues associated with data dissemination, which include content scheduling, data structures and format of dissemination cycle, and concurrency control techniques. There has been much work examining how to determine what data to disseminate based on user needs. In addition, there is a lot of work that addresses the structure of the dissemination cycle. Tsakiridis et al and Wang et al study how to improve the dissemination network performance through optimizing data allocation in [1] and [2]. In the field of a query processing in multi-cell wireless environment, Jayaputera et al discuss the issues of retrieving the data for location dependent query in [3]. In addition, Kwangjin investigates how best to disseminate data in the network in [4]. A more comprehensive work of how to access the disseminated data and adapt it to the wireless and mobile computing environments is discussed in [5], [6], and [7].

Data dissemination environments and applications require special architectural models. In addition, their usage must be incorporated into the more traditional mobile computing environment for both mobile transactions and queries, and that where centralized and distributed database access occur. Moreover, it is important to adapt the architectural model in the telecommunication network. The wireless infrastructure of the future must support all means of data access not just push or pull, wired or wireless, retrieval or update.

In earlier work, we showed a redesign of the complete wireless data access infrastructure, which supports both push and pull access to data [8]. Specifically, we outlined a design of architecture, which is flexible and scalable. This architecture is for the multi-channel dissemination network, which works in conjunction with the general telecommunication network.

In this paper, we present a concurrency control technique suitable for the multi-channel dissemination-based architectural model. This approach facilitates the update of data read from dissemination network. This approach is a combination of processing algorithms, concurrency control algorithms, and dissemination structure. In this paper, we report on the results of a performance study, which examines the overhead associated with it. The study shows the superiority of the proposed technique by reducing the overhead and improving data access.

The paper is presented as follow. In section 2, we discuss the proposed technique for the multi-channel dissemination architectural model that fits the general telecommunication network. In section 3, a performance study for the system overhead is shown. Finally, the conclusions and future work are presented in section 4.

## 2. THE PROPOSED TECHNIQUE FOR THE MULTI-CHANNEL DISSEMINATION ARCHITECTURE

Wireless applications can not work independently and separately from traditional database transactions. It is expected in the foreseeable future, all types of data access occurring simultaneously from traditional centralized and distributed sources (i.e. online access from terminals, batch transactions, internet applications, etc.) and from wireless applications. This includes both mobile transactions/queries and dissemination applications. An architecture which facilitates the seamless execution of these diverse types of access is needed. Although each of these three types of access (wired, wireless, dissemination) have been investigated separately, little work has examined how they fit together or projected in the telecommunication network.

The dissemination architecture easily supports all types of access. To facilitate understanding and appreciation of this encompassing architecture, we show the different components into the following functional units.

- *Contents Providers* (CP)s provide the data to be read and updated by all users.

- *Dissemination Operators* (DO)s are responsible for the actual *push* of data. The CPs feed the DOs with the needed disseminated data.

- *Mobile Support Stations* (MSS)s are the traditional platforms which support a two-way wireless communication with the wireless users.

- *Dissemination Controller* (DC) manages data to be disseminated from the CPs and MSSs to the DOs. To accomplish this objective, it must perform several functions. First it must manage channels by determining the content and usage of channels. One or more controller may exist. It is assumed that each controller may be responsible for many DOs and thus all of their channels. The DC maintains user profiles which are used to determine the content of dissemination channels. In addition, mobile users may predefine their movement intentions or request specific data. This information is also stored and used to determine the dissemination content.

- *Wireless Clients* (WC)s are users with devices (e.g. PDA, cell phones, etc.) through which they can access the desired information. The WCs need to tune in to the desired channel to obtain the data.

These functional units of the dissemination-based architecture are illustrated in Figure 1. The DC manages data between the CPs and MSSs, in the one hand, and the DOs in the other hand. In fact, each DO may actually disseminate over multiple channels. Each WC may access one or more channels from one or more MSSs.

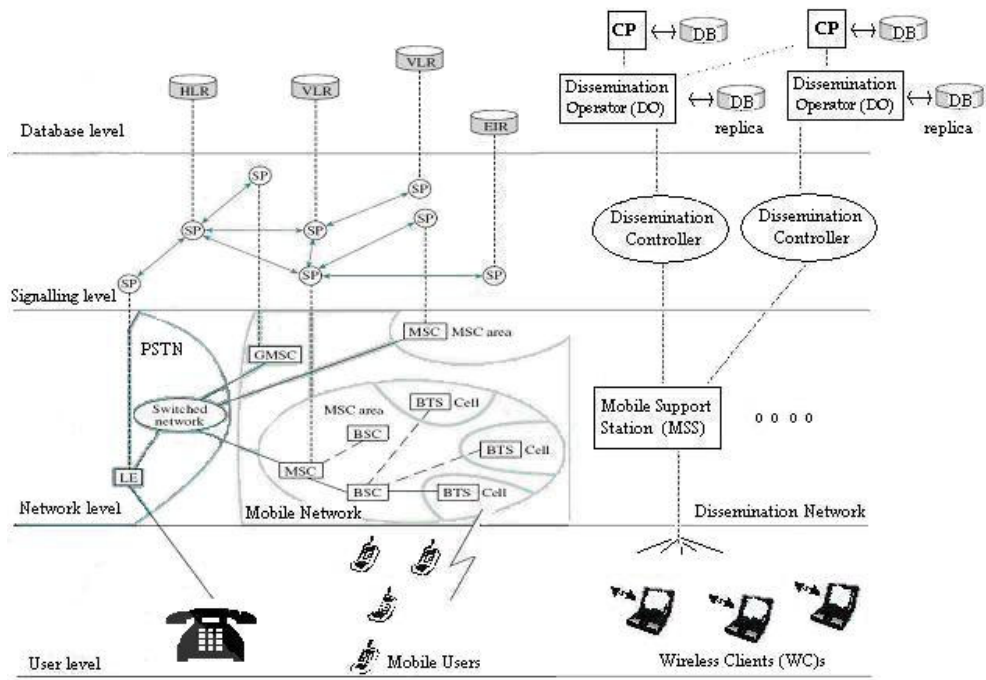

Figure 1. The proposed dissemination architecture within the telecommunication network

When the content of dissemination is changed, the controller requests new data from the CP via read transactions and sends the data to DOs for dissemination. As shown in the figure, the WC can only read the disseminated data. However, if WC wishes to perform updates, a separate more traditional two way wireless architectural component with MSSs must exist. For simplicity, this figure shows the general usage of a conventional cellular structure as described in [9], which may exist concurrently with the multi-channel dissemination architecture. The functional units of the dissemination architecture are divided in the four levels of the cellular network as follow.

- *The Database level.* In this level, *Contents Providers* (CP)s supply the data to be disseminated which they maintain and manage. These may be legacy systems which support more conventional relational databases and SQL access. The CPs feed the *Dissemination Operators* (DO)s with the desired contents to disseminate through the MSSs. The DOs maintain *read-only* databases.

- *The Signaling level.* In this level, one or more *Dissemination controller (DC)* is required to specify what and when data items are put in the wireless dissemination channels. Clients need to tune in to the correct channel(s) (channel identification) and find the desired data (through a data indexing component). It is expected that the future wireless environment will consist of many controllers and many DOs, just as many wireless operators coexist today.

- *The Network level.* In this level, *MSSs* provide the services and infrastructure needed to support the normal two-way wireless phone/data usage.

- *The User level.* In this level, *WCs* read the disseminated data from their devices. Also, clients will be able to interface directly with the the dissemination controller via specific predefined features provided directly or through internet connections. These will allow the user to predefine data needs prior to travel and to sign up for desired services.

The CPs and the WCs will subscribe to the services of the dissemination operators. The CPs agree to provide the data if the DOs agree to push it under predetermined conditions, such as the frequency. The WCs require to sign up for the data dissemination services they need and each WC provides the CPs some information to create a profile and may even indicate normal movement patterns which can be used to help determine the contents of the dissemination channels.

In actuality, clients may not even need to subscribe to a given service. The content provider may be willing to pay the entire cost for pushing his data in exchange for the advertising he receives as a result. For example, major hotel chains may wish to have information concerning their local hotels disseminated from dissemination operators in the area. They want as many users as possible to read this information, so they request that any clients be able to access their data free of charge.

Since there may be different data formats and different user requests for data we may use different *channels*. Each channel can be used to support different dissemination needs for a set of clients. For instance, one channel may carry flight information, another channel may carry city attractions for a group of users at the airport. Also, these channels may support different formats for data, such as self-indexed and full-indexed. The basic idea is to have one or several CPs to feed one or several DOs. This makes it possible to support different user profiles requests for different group of users. Data may be accessed thorough various means. Traditional Transactions access data at CP(s) using traditional DBMS facilities such as SQL. A Read Transaction is a Traditional Transaction, which only performs read operations. With Mobile Transactions (MT)s, WC accesses data at the CP(s). Dissemination Transactions (DT)s are special transactions that access the data at CP(s) and then disseminate this data from DO(s). Write Transactions (WT)s are transactions to update data, which previously read from a dissemination network. Users can then perform read access to the data through a Traditional Transactions, Mobile Transactions, or Dissemination Transactions. Users can perform read/write access to the data through a TT, MT, or DT/WT combined. The architecture decouples the disseminate function from that of providing the data to disseminate. Moreover, the architecture allows the WCs to update the content of the dissemination. There is no interrupt for the dissemination of data until the end of its dissemination cycle.

The consistency checking is insured through using a Multi-Channel Dissemination (MCD) technique. The MCD is the concurrency control of data items at the CPs and the DOs. The basic idea of the MCD is to have several channels in the dissemination environment. Each WC tunes in to the designated channel(s) according to the WC's profile. Each Mobile Transaction (MT) has a Read Set (RS), which contains all data items the MT needs to read from the DT. When an WC sees a data item that is on the RS for the MT, the WC needs to read the data item from the dissemination. If the MT performs a read only operation with all data coming from one dissemination cycle, the MT may commit locally at the WC without contacting the CP(s). When this is not the case, a transaction has to be sent via a backchannel to the CP(s) for validation. When executing at the CP, traditional locking techniques are used. MCD technique insures consistency of dissemination by placing additional control information at each DT. The contents of the DT are, (channel number, dissemination cycle, index, items).

When the DO dissemination program starts, it sends the channel number, current dissemination cycle, index, and the list of data items to the listening WCs. The channel number determines which channel the WC needs to tune in to get the needed data items. The index determines the location of each data item in the DT. In addition, the index contains some control information about each data item in DT (e.g. last time data item disseminated, updated or not since last pushed, etc.). The WC uses the control information to determine where to read his RS data items in the DT. The DO never gets interrupted when it starts disseminating DT. For instance, when a write transaction at CP occurs, the CP schedules those updated items to the next dissemination cycle. The CP changes the control information of those updated items to reflect the next (disseminate and update) cycles for items.

At the start of next dissemination cycle, the WC, again, uses the control information in the index to find out whether data items are disseminated again, or to check if there is any new data items in his RS. This control information is very useful for a subset dissemination of database. When the WC is done with all his read and write data and operations and the WC sees that all data items read from the DTs are consistent (i.e. which occurs when all $updatedcycle_{DT}(x_i)$ is the same as the $updatedcycle_{MT}(x_i)$ for all read data items $x_i$ at RS of MT), then all data items of ReadSet (RS) and WriteSet (WS) are sent to CPs for commit. If the MT is read-only and consistent with the data of the DT, then the MT may commit at current cycle without contacting the CPs. Once the CP receives the request for the update from clients, the CP then assigns an *S* lock of the RS of WT and an *X* lock for the WS of WT. The *X* lock allows only one transaction at a time to access the data item (i.e. mutually exclusion access) while the *S* lock ensures no other concurrent transaction is going to update any data in the RS of the WT. If all needed *S* and *X* locks are acquired, the WT is allowed to proceed and update the WS at the CP.

When the WT finishes, all *S* and *X* locks get released and a confirmation to the WC is sent in a mutually exclusive way. Otherwise (i.e. when RS is inconsistent with CP which happens when there is at least one data item value read at the MT from the DT is different than the CP), the CP informs the MT of the inconsistent values through the backchannel, and ask the WC to re-read all inconsistent data items again from the dissemination. Moreover, the designated CP prioritizes the dissemination of the inconsistent data items and push them again in the next cycle. This way, the WC does not need to abort the whole transaction and restart reading all data items from the dissemination network. Instead, the WC needs only to read the inconsistent data items in the RS. This improves the performance of algorithm and reduces the overhead associated with frequent aborts.

## 3. PERFORMANCE STUDY

In this section, we report the results of a performance study to examine the overhead of the system. We compare the performance of *MCD* to *fresh*, *nxn*, and *Perfect* approaches. While we examined the *MCD* approach in the last section, we will shed the light on *fresh* and *nxn* techniques in the next paragraph. The *Perfect* approach assumes there is no overhead associated to disseminate data. This provides the best case to compare other approaches.

The *fresh* and *nxn* techniques are used for concurrency control in the data dissemination network in the context of a transaction as presented in [10] and [11], respectively. The former uses Update-First with Order to ensure the concurrency in the dissemination with *fresh* data, whereas the latter approach uses a matrix of size *nxn* for concurrency control where *n* is the number of data items in the database. In the *nxn* approach, the user can both read and update the data, while in the *fresh* approach, the user is *read-only*. When an update occurs at the database provider of dissemination network, the new updated item is uploaded in the next subsequent cycle in the *nxn* approach, whereas it interrupts pushing data to upload the updated new item in the *fresh* approach. As a part of the simulation work, we show the improvements of MCD using several parameters, workloads, and experiments. The simulators used are implemented using C++ language. The total amount of code is over 1700 LOC (Lines Of Code). We used both black and white box testing to ensure the accuracy and logic of the programming.

In running the simulation, the set of data items loaded at the server are predetermined and run at the start of executing the simulation program. Since we use the approach of predetermined items, the overhead of uploading them is omitted. The predetermined items are selected randomly according to the running experiment. Nonetheless, the WC always gets the needed items at some point in the dissemination cycles and channels. Since the data items of the WC are predetermined, there is no I/O operation. The updates needed to items in the CP database are conducted by some local update transactions at the CP.

Moreover, parts of the databases of the CP(s) are disseminated in one cycle. In this view, the DT contains some database contents of the CPs (i.e. what is uploaded at DO). Based on the DC, different subset of database is disseminated to the WC in cyclical way. The dynamic behavior of the simulation occurs through changing the sizes of operations, MTs, and DT.

The setting is applied to all running algorithms. There are several wireless dissemination channels. Clients need to tune in to the correct channel(s) using channel identification and find the desired data through a data-indexing component. The query generators at the WCs are responsible to determine the needed items to be read by for all MTs at the start of dissemination. The processor and scheduler at WCs determine when/where to listen to DTs according to the way each algorithm functions.

In the following, we show the results of the simulation. They are based on the Response Time (RT), Power Consumption (PC), and Space Overhead (SO) for all the experiments.

- *The effect of the checking cost on response time.* This is shown in Figure 2. It depicts the cost of determining whether the disseminated item is in the read set of the WC. This *decision* cost increases the response time of the *fresh* approach since this approach does not utilize the use of indexing. This makes *fresh* approach check every item in the dissemination channel(s) to see whether it is needed or not (i.e. in the read set). For the other three approaches, the checking cost is steady since the mobile unit calculates exactly when and where it should read the data from the channel(s).

- *The effect of the item size change.* This experiment shows the impact of changing the item size at the DO to see the effect on the performance of the system.
    - For the results of the Response Time (RT) and the Power Consumption (PC), as the size of the item grows, much time and power are needed to read the item from the dissemination. For *nxn* and *MCD*, the control information overhead stays the same despite changing the size of item. For *fresh* approach, there is no control information and the time and power needed are for reading the item. However, as the size of item changes, the *fresh* approach interrupts the DT more to upload the new item and causes overhead and power consumption as depicted in Figure 3.
    - For the Space Overhead (SO), the *nxn* approach consumes less bandwidth for the control information than *fresh* approach when the item size gets bigger. This makes the effect of the control information in *nxn* approach less noticeable for some applications, such as multimedia. However, the *MCD* outperforms both *nxn* and *fresh* techniques as shown in Figure 4.

- *The effect of the item order*. This experiment shows the impact of placing the needed item in the dissemination. This shows how much it costs to listen to the dissemination channel when the WC does not utilize the sleeping mode. The effect is seen in *fresh* algorithm since the WC has to listen to DT at all times. The *MCD* shows a good result, which is close to that of *perfect*, while *fresh* lags behind. Since *fresh* technique causes the client to tune to DT all the time, its overhead is proportional to the place of item. The *MCD* and the *nxn* techniques use index to tune to the dissemination channel and, therefore, the listening cost of tuning is already reflected in the index. Nonetheless, *nxn* approach uses much overhead due to the large size of the *nxn* as shown in Figure 5.

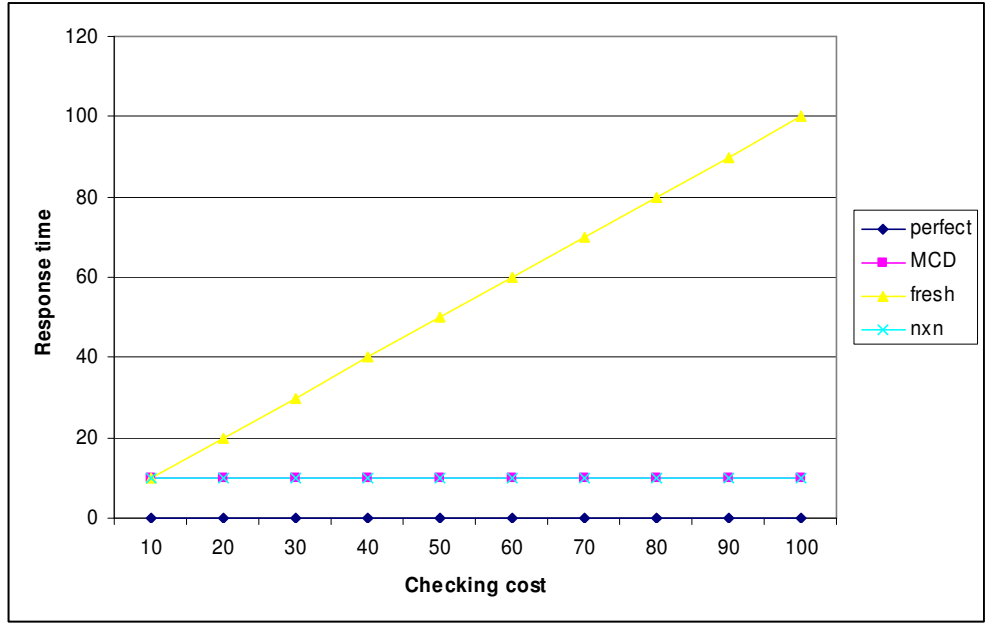

**Figure 2.** The effect of the checking cost on the response time

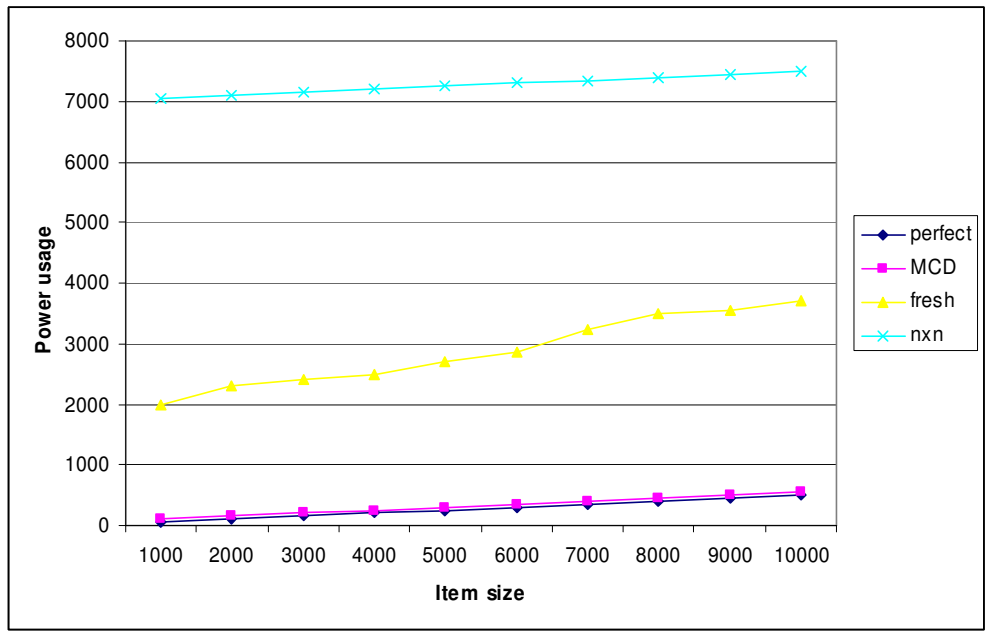

**Figure 3.** The effect of the item size change on the power usage

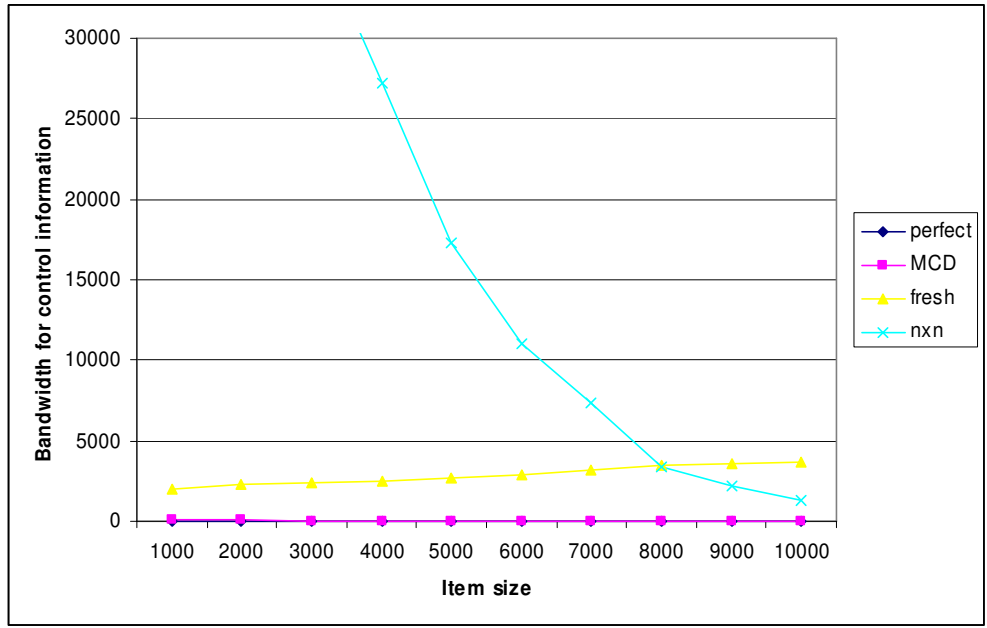

**Figure 4.** The effect of the item size change on the space overhead

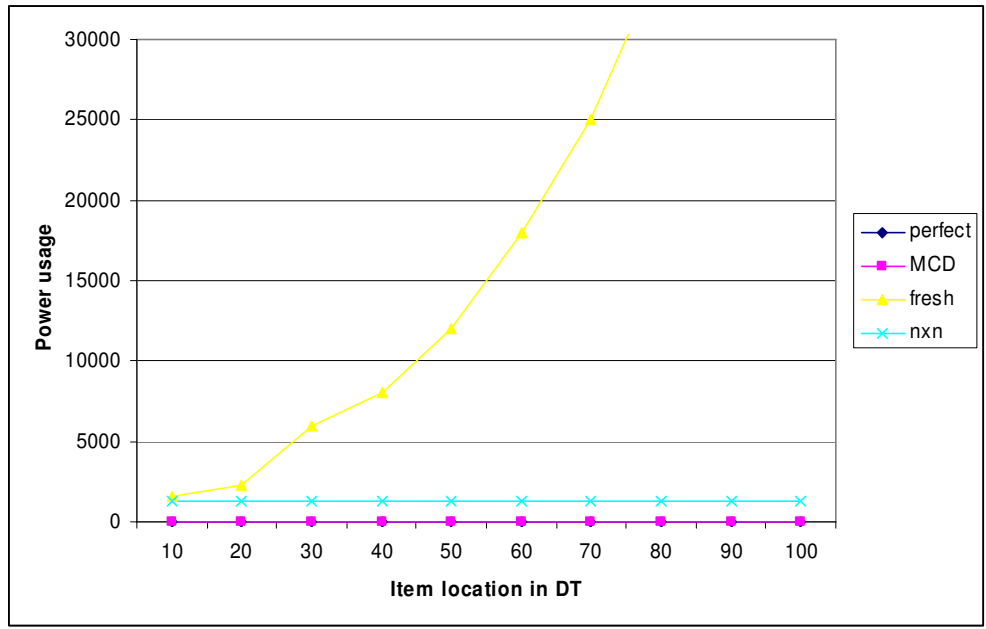

**Figure 5.** The effect of the listening cost change on the power usage

139

## 4. CONCLUSIONS AND FUTURE WORK

This work shows the efficiency and power saving for a proposed technique, MCD, which insures the concurrency of data items in the multi-channel dissemination network. The MCD approach is compared to the *fresh* and *nxn* approaches. The simulation shows how much *fresh* approach consumes a lot of processing power due to the continuous tuning to the dissemination channel. In addition, the simulation indicates that *fresh* approach does not perform well when the update rate is high. However, *fresh* approach works fine in low update environment, such as the weather dissemination environment. The *nxn* approach may fit very well in some specific environments where the size of items is very large and the number of items is small as the simulation findings show. Such an example is the wireless multimedia environment where the size of data items tends to be large. Finally, the simulation study shows how the response time is improved, the space overhead is reduced, and the power is saved when MCD is used to ensure the concurrency control in the dissemination environment.

As future work, it may be appropriate to investigate how the proposed architecture works with the use of location dependent data and caching. In addition, it is critical to look at the effect of the user movement and the management issues involved to maintain users information. Finally, at the current time, we are developing a wireless network using a *Wireless Fidelity* Wi-Fi (802.11n) and implementing various dissemination methods to check the actual performance.